\documentclass[letterpaper]{article}

\newcommand{\argmax}{\operatornamewithlimits{argmax}}

\usepackage{graphicx}
\graphicspath{{./Fig/}}
\usepackage{color}
\usepackage{placeins}
\usepackage{float}
\usepackage{tabularx,colortbl}
\usepackage{amsmath}
\usepackage{amsfonts}
\usepackage{caption}
\usepackage{subcaption}
\usepackage[ruled,vlined]{algorithm2e}
\usepackage{cite}
\usepackage{natbib}
\usepackage{epstopdf}
\newcommand{\comment}[1]{}

\usepackage{authblk}
\usepackage[margin=1.0in]{geometry}

\begin{document}

\title{Iterative interferometry-based method for picking microseismic events}
\author[1]{Naveed Iqbal\thanks{naveediqbal@kfupm.edu.sa}}
\author[1]{Abdullatif A. Al-Shuhail\thanks{ashuhail@kfupm.edu.sa}}
\author[1]{SanLinn I. Kaka\thanks{skaka@kfupm.edu.sa}}
\author[2]{Entao Liu\thanks{eliu@gatech.edu}}
\author[2]{Anupama Govinda Raj\thanks{agr6@gatech.edu}}
\author[2]{James H. McClellan\thanks{jim.mcclellan@gatech.edu}}
\affil[1]{CeGP at KFUPM}
\affil[2]{CeGP at Georgia Institute of Technology}

\renewcommand\Authands{, and }


%Telephone: (800) 555--1212, Fax: (888) 555--1212}


\maketitle
\begin{abstract}
Continuous microseismic monitoring of hydraulic fracturing is commonly used in many engineering, environmental, mining, and petroleum applications. Microseismic signals recorded at the surface, suffer from excessive noise that complicates first-break picking and subsequent data processing and analysis. This study presents a new first-break picking algorithm that employs concepts from seismic interferometry and time-frequency (TF) analysis. The algorithm first uses a TF plot to manually pick a reference first-break and then  iterates the steps of cross-correlation, alignment, and stacking to enhance the signal-to-noise ratio of the relative first breaks. The reference first-break is subsequently used to calculate final first breaks from the relative ones. Testing on synthetic and real data sets at high levels of additive noise shows that the algorithm enhances the first-break picking considerably. Furthermore, results show that only two iterations are needed to converge to the true first breaks. Indeed, iterating more can have detrimental effects on the algorithm due to increasing correlation of random noise.\end{abstract}

\newpage
\begin{keywords}
cross correlation, first-break, time-frequency representation, P- and S-arrivals.
\end{keywords}

\section{Introduction}

Renewed interest in microseismic monitoring of hydraulic fracturing for the development of unconventional reservoirs has led the petroleum industry to invest in microsesismic technology. The technology involves installing downhole and surface sensors to detect and localize microseismic events in order to image fracturing associated with fluid injection or extraction. The microseismic fracture image is used to understand the extent of fracture network. However, microseismic events are known to have very low signal-to-noise ratio (SNR) making it difficult to detect and accurately locate microseismic events.  Many researchers have proposed seismic interferometry to increase the SNR of seismic data \citep{Sni2004,Wapvand2008,Sch2009,WapDraAO2010,WapSloAO2010}. Recent application of seismic interferometry in first-break enhancement through cross-correlation, summation, and convolution showed promising results when applied on seismic signals generated by an active source, e.g., \citep{MalBhaAO2011,AlAAldAO2012,BhaWanAO2013,AlHHanAO2014}.

\cite{XiaLuoAO09} used seismic interferometry to locate microseismic events by cross-correlating the direct P and S arrivals from repeated sources. They tested their approach using an elastic model and found that the repeated sources enhance arrivals after stacking. \cite{SonKulAO2010} proposed an array-based waveform correlation method to enhance the detectability of microseismic events. They used a transformed spectrogram to identify the arrivals and found an improvement over an array-stacked short-time average/long-time average (STA/LTA) approach. There are various versions of STA/LTA method, which are based on energy, amplitude or entropy functions \citep{Wong2009,Xiantai2011}. There are other similar methods, e.g., multi-window method \citep{Chen2005} and modified energy ratio (MER) \citep{Wong2009}. \cite{Mousa2011} discussed a method based on the digital image segmentation. \cite{AlSKakAO2013} proposed a workflow to enhance microseismic events and reported that the last step in their workflow (i.e., convolving the enhanced and raw records) seems responsible for leaking noise from the raw records into the enhanced data. This problem is eliminated  by \cite{Iqbal2016b} using singular-value decomposition in place of convolution.  

Among the  existing methods, STA/LTA method is the most widely used in earthquake seismology. The STA/LTA method processes the signals in  two moving windows (long and short). The ratio of the average energies in the windows are calculated. In case of noise-free seismograms, the maximum value of numerical derivative of the ratio is close to the first-arrival time. In the MER method, the concept of STA/LTA is modified in such a way that two windows are of equal length and move adjacent to each other. The peak of the cubic power of  energies ratio in two windows   is close to a first-break. The other technique to detect the first breaks includes interferometry approach where cross-correlations of all the distinct pair of traces are carried out. Then, the cross-correlations are aligned to a specific instant and summed up. The resulting stacked cross-correlation is considered as a filter to denoise the noisy traces by convolving the stacked cross-correlation with the traces.  \   

In this study, we propose a first-break picking approach based on seismic interferometry.
% This approach does not involve convolution of the enhanced and raw traces ensuring that the raw and enhanced records do not mix. Moreover,
The proposed approach applies the basic steps of seismic interferometry (cross-correlation and summation) iteratively in order to minimize the sum of squared errors (SSE) between successive iterations. We use both synthetic and real microseismic data to demonstrate that this new approach significantly enhances first breaks, thus making it easy to pick event arrival times. Consequently, the enhanced signals with accurate picking of first breaks will likely improve microseismic event localization over the reservoir.

\section{Proposed Method}
%There are seven steps to our method: manual  picking, cross-correlating, picking the timing of the maximum value, aligning the maxima, stacking, iterating the entire workflow and calculating the enhanced first breaks.

The proposed method has two phases: reliable/accurate first-break estimation on a reference trace and estimation of the relative time delays from the reference trace to all other traces with an enhanced cross correlation function (CCF).
Assume there are $M$ raw microseismic traces, $x_m[n]$ for $m=1,\ldots,M$ of length $L_T$ and sampling interval $\Delta t$. With $M$ traces, the number of unique CCFs is $Q=(M/2)(M-1)$.

 The success of the proposed method depends on the correct picking of a reference first-break on at least one trace; therefore, we have given special attention to this step.  Consequently, we use the time-frequency  contents of the traces simultaneously in the time-frequency domain, which will make the process of first-break manual picking more reliable.  Time-frequency representation is not a new concept and many high resolution time-frequency decompositions have been developed. The short-time Fourier transform (STFT) and the continuous wavelet transform are well-known time-frequency transforms that can recover the signal contents if they do not overlap in the time-frequency domain \citep{Diallo2005,Kulesh2007,Roueff2004}. There are other transforms such as the empirical mode decomposition \citep{Huang1998}, synchrosqueezing transform \citep{Daubechies2011}, matching pursuit \citep{Mallat1993} and basis pursuit \citep{Bonar2010,VeraRodriguez2012,Zhang2011}, to name a few.  The time-frequency representation has been used for  travel-time picking in the past \citep{Saragiotis2013,Zhang2015,Herrera2015}. In this work, we have tested various time-frequency transforms but find spectrogram method to be most effective. Hence, we use the  spectrogram method proposed by \cite{AugFla1995}  to pick a reference trace $x_r[n]$ whose first-break $n_r$ is the most clear in the time-frequency domain.

For Spectrogram, Fast Fourier Transform (FFT) is applied on subset of data points $(N)$. This subset of data is selected using a window $\mathbf w=[w_0, w_1,\cdots,w_{L_w}]$. First, FFT is computed for data points of length $L_w$, then the window in moved $h$ data points and again FFT is calculated. Here, $h$ is called as step size. This procedure is repeated until the window covers the last $L_w$ data points. Mathematically,
\begin{eqnarray}
Y(p,q)=\sum_{j=0}^{L_{w}-1}y_{j+qh}w_{j}\exp\left({-i2\pi jp\over L_{w}}\right),\ p=0,1,\cdots,L_{w}-1,\ q=0,1,2,\cdots,{N-L_w \over h}
\end{eqnarray}

The spectrogram is calculated as$|Y(l,n)|^{2}$.
We have used the Hamming window of length $(L_w)$  quarter of the trace length and the overlap is $L_w - 1$, i.e., $h=1$ (these parameters are optimized using simulations).

The reference time index $n_r$ will be utilized to calculate the absolute timings of the enhanced first breaks after the second step which involves multiple cross correlations. The units of $n_r$ are number of samples, which can be converted to time using the sampling interval, i.e.,  $t_r = (n_r-1) \Delta t$. Another step in the method is to accurately find the relative time delays of all other traces to the reference trace. Then, the first-break for an arbitrary trace is the sum of $t_r$ and the relative delay to the reference trace.
See Algorithm \ref{alg:iter} for a detailed description of the proposed scheme.
\\ \emph{Notation:}
For two distinct traces, the CCF is defined as
\begin{equation}
        \Phi_{lm}[\tau] = x_l\otimes x_m =\sum_n  x_l[n]  x_m[n+\tau],
\end{equation}
where $\otimes$ denotes the convolution operator and $\tau$ is the lag index, and $-L_{T}\le \tau\le L_{T}$, ($L_T$ denotes the length of the traces. i.e., total number of samples). 

\begin{algorithm}
  \caption{Iterative first breaks picking method based on CCF}
  \label{alg:iter}
  %\SetAlgoLined
  %\donotprintsemicolon
  \SetAlFnt{\small}
  \SetKwInOut{KwInit}{Initialization}
  \AlFnt
  \KwIn{Raw microseismic traces $x_m[n]$, $m=1,\ldots,M$.}
  \KwInit{ Manually pick the first-break $n_r$ of the reference trace. \\
}
  \begin{itemize}
        \item [Step 0:] Set iteration counter $i=0$. \\
            Compute cross-correlation between trace $l$ and $m$, $\Phi^i_{lm}[\tau]$\\
            $\Phi^i_{lm}[\tau] = x_l\otimes x_m $ for $l<m$, $m=2,\ldots,M$ and $-L_{T}\le \tau\le L_{T}$ \\

\item [Step 1:] Pick relative time delays between the reference trace $r$ and all other traces
\begin{equation*}
        \tau^i_{rm} = \begin{cases}
          {\displaystyle \argmax_\tau \Phi^i_{rm}[\tau]} & r<m\\
          0 & r=m\\
          {\displaystyle -\argmax_\tau \Phi^i_{rm}[\tau]} & r>m  \end{cases}
     \end{equation*}

      \KwOut{first breaks for all traces: $n^i_m= n_r+\tau^i_{rm}$, $m=1,\ldots,M$. }

\item[Step 2:] If $i\geq1$, compute iterative sum of squared errors, ISSE$(i)$ via
\begin{equation*}
   \text{ISSE}(i) =  \sum\limits_{m=1}^M \left(n^i_m - n^{i-1}_m \right)^2
     =  \sum\limits_{m=1}^M \left(\tau^i_{rm} - \tau^{i-1}_{rm} \right)^2
     \end{equation*}
Terminate the iteration if ISSE$(i)$ increases.

\item [Step 3:]
 Align the maxima and stack the CCFs
\begin{equation*}
\Phi^{i}_s[\tau] = \frac{1}{Q} \sum_{l<m} \Phi^{i}_{lm}[\tau-\tau^{i}_{lm}], \mbox{ where } \tau^{i}_{lm}=\argmax_\tau \Phi^{i}_{lm}[\tau]
\end{equation*}

\item[Step 4:]
Update the CCFs via convolution and truncation\begin{align*}
\hat{\Phi}^{i+1}_{lm}[\tau] &= \Phi^{i}_{lm}\otimes \Phi^{i}_s[\tau]\\
\Phi^{i+1}_{lm}[\tau] &\gets \hat{\Phi}^{i+1}_{lm} [\tau]T[\tau],
\end{align*}
where $-L_{T}\le \tau\le L_{T}$ and $T[\tau]$ is a truncation window such that
\begin{equation*}
        T[\tau]=\left\{
        \begin{array}{ll}
                1 & \mbox{ for } -N_T\le \tau \le N_T\\
                0 & \mbox{ otherwise}
        \end{array}\right.
\end{equation*}
\item[Step 5:]  $i\gets i+1$.  Go to Step 1.

  \end{itemize}

\end{algorithm}
  The algorithm use $\displaystyle \tau_{lm} = \argmax_\tau \{\Phi_{lm}[\tau]\}$ as the initial picks of the relative delays. These delays  enable CCF stacking (step 3 of the Algorithm \ref{alg:iter}) which is a key step in this proposed scheme, where we iteratively perform cross-correlation of the CCFs to mitigate the negative impact of the high noise on the first-break picking. However, in the sequel we will also see that over iterating this scheme will ultimately harm the results.

It is worth mentioning here that in case of both P- and S-arrivals, the algorithm is applied sequentially to find both arrivals. Since S-waves have relatively high amplitudes than P-waves, application of the algorithm in the first place  leads us to estimate  S-arrivals. After getting the S-arrivals ($S_{m,arrival}$),  the P-arrivals  are found by again applying the algorithm on the data set with S-waves removed. This is done by taking the part of data up to the $S_{m,arrival}$, i.e,
\begin{align}\label{win1}
 x_m[n]= x_m[n] p_m[n], \ m=1,\ldots,M
\end{align}
where
\begin{equation}\label{win2}
 p_m[n]= \left\{
        \begin{array}{ll}
                1 & \mbox{ for } 1\le n \le S_{m,arrival}\\
                0 & \mbox{ otherwise}
        \end{array}\right.
\end{equation}
Accordingly, the reference trace will be  used to pick S-arrival reference in the first place and P-arrival reference in the second place.

\section{Results}
This section outlines the performance of the proposed cross-correlation based method for the cases of extremely noisy synthetic and real data records.

\subsection{Test on synthetic microseismic data}
\begin{figure}[htbp]
\begin{center}
\includegraphics[width=0.8\textwidth]{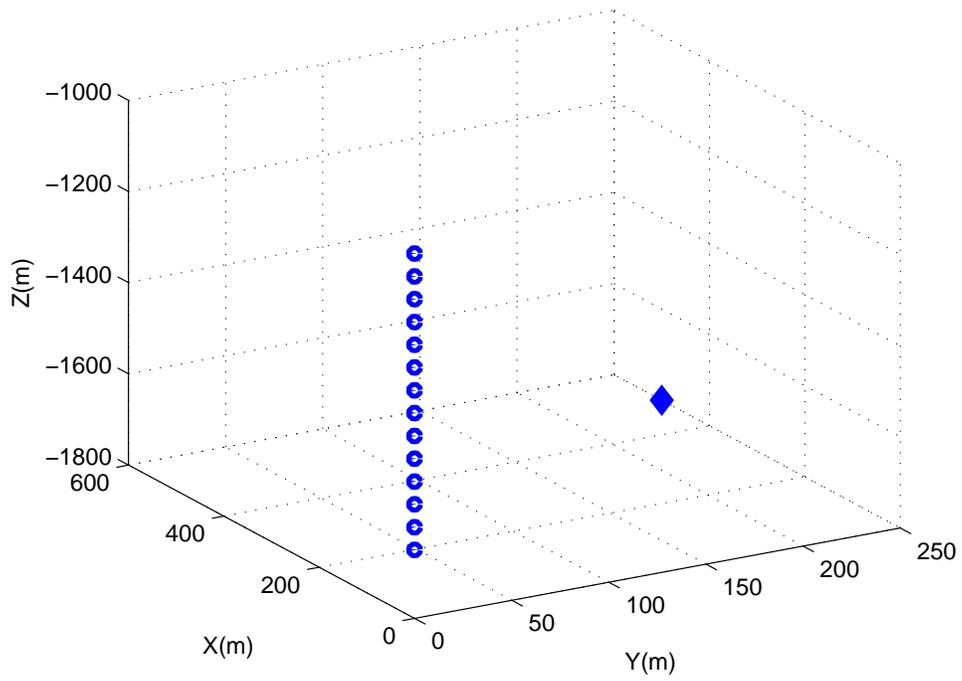}
\caption{Geometry of microseismic source at (500,250,$-$1800)\,m and borehole receivers at (0,0, $z_m$) used to generate synthetic data. Circles represent borehole receivers, whereas, diamond represents the microseismic source. }
\label{fig:setup}
\end{center}
\end{figure}

First, we apply our work flow on  a synthetic microseismic record generated using the following parameters:
\begin{itemize}
        \item Receivers are located in a vertical borehole at  $x_m=0$, $y_m=0$, $z_m=-1000-(m-1)\Delta z$ for $m=1,\ldots,M$, where $M=14$ and $\Delta z=50$\,m as depicted in Figure \ref{fig:setup}.
        \item Constant medium velocity of $2000$ m/s.
        \item Source wavelet is a normalized minimum-phase Berlage wavelet \citep{Ald1990}
              \begin{equation}
                W(t) = At^n e^{-\alpha t} \cos(2\pi f t +\phi),
              \end{equation}
              with $A=1$, $n=0.001$, $\alpha=15$, $f=5$\,Hz, and $\phi=-\pi/2$. This wavelet is frequently used to represent a seismic source in the literature (e.g., \cite{Poiata2016}).   
    \item Sampling interval is $\Delta t=1$ ms with 1001 samples per trace.
    \item Source coordinates are $x_s=500$\,m, $y_s=250$\,m, and $z_s=-1800$\,m.
        \end{itemize}

         \begin{figure*}
        \centering
        \begin{subfigure}[b]{0.475\textwidth}
            \centering
            \includegraphics[width=1\textwidth]{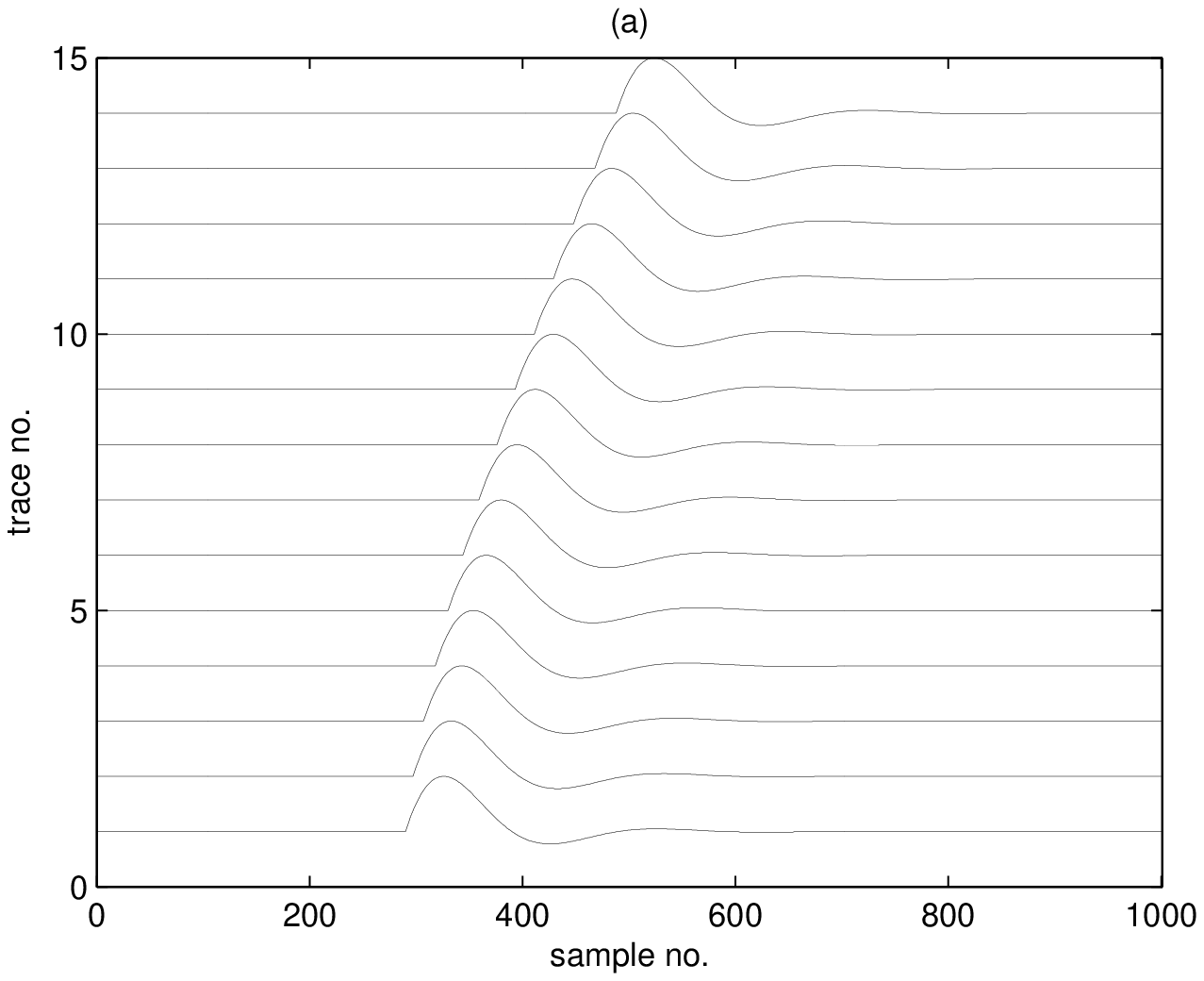}
             \end{subfigure}
        \hfill
        \begin{subfigure}[b]{0.475\textwidth}
            \centering
            \includegraphics[width=1\textwidth]{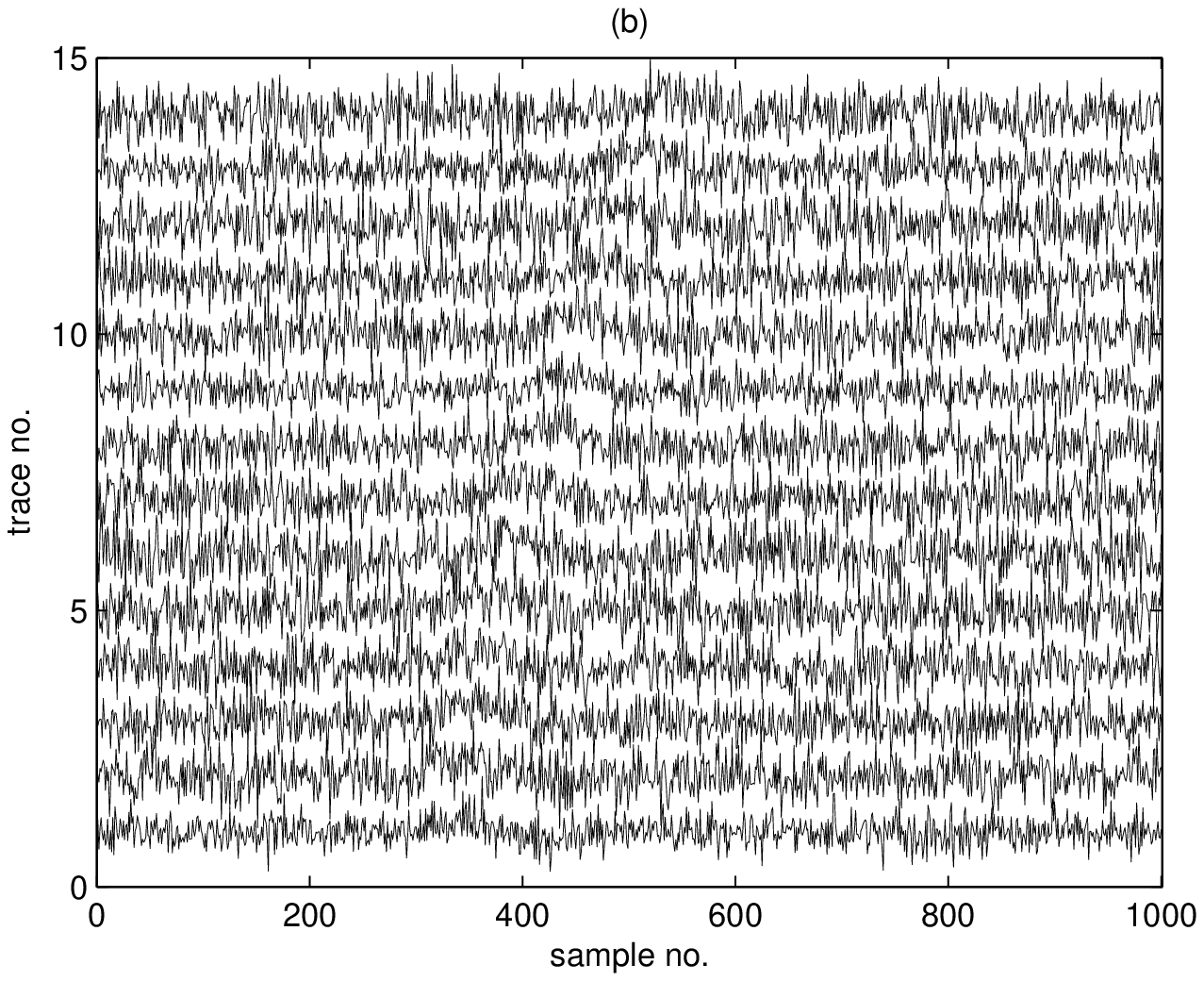}
            \end{subfigure}
        \caption[ The average and standard deviation of critical parameters ]
        {\small Synthetic data set. (a) noiseless traces, and (b) noisy traces  SNR = $-$12\,dB.}
        \label{fig:NNtraces}
    \end{figure*}

\begin{figure}[htbp]
\begin{center}
\includegraphics[width=0.75\textwidth]{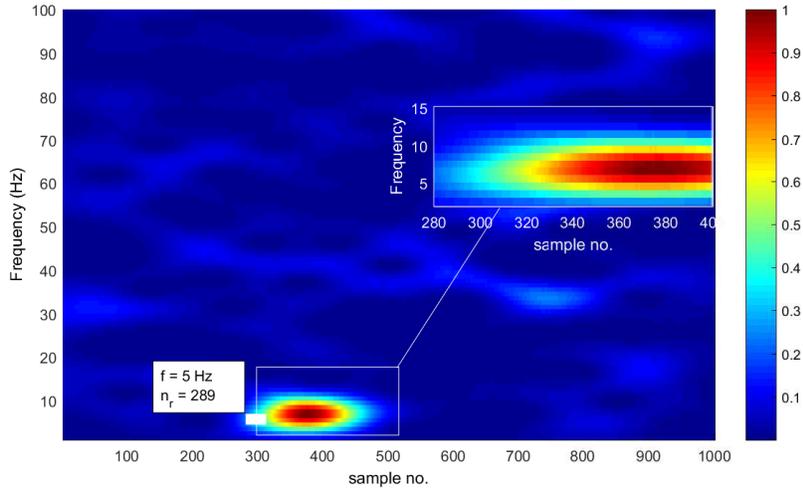}
\caption{
Spectrogram of the first trace and picked first-break at $n_r = 289$ (indicated by the small square);
the inset shows a zoomed view into the region near the first-break. Window length is 250; FFT length is 250 as well.}
\label{fig:TF}
\end{center}
\end{figure}

        Figure \ref{fig:NNtraces}a shows the traces before adding noise in order to illustrate the correct first breaks. Figure \ref{fig:NNtraces}b shows the traces after adding noise and trace normalization. The additive white Gaussian noise has zero mean and variance $\sigma_n^2$, and the noise less signal has variance $\sigma_s^2$. The signal-to-noise ratio
        \begin{equation}\label{PSNR}
\text{SNR}=10\log_{10}\left(\frac{\sigma_s^2}{\sigma_n^2}\right),
\end{equation}
is set to $-$12\,dB for the synthetic data and the corresponding peak-signal-to-noise ratio
 \begin{equation}\label{PSNR}
\text{PSNR}=10\log_{10}\left(\frac{P^2}{\sigma^2}\right),
\end{equation}
is 1\,dB, where $P$ is the peak value of the known wavelet. When the wavelet duration is short, the PSNR is more intuitive than SNR, because its value is not affected by the signal length. At this PSNR level, automatic first-break picking from the noisy records is challenging. Figure \ref{fig:TF} depicts the spectrogram or time-frequency transform (TFT) of trace 1 of the synthetic data.
 The time-frequency processing uses a Hamming window whose length is $L_T/4$, where $L_T$ is the total length of a trace. The time-frequency plot of trace 1 in Figure \ref{fig:TF} shows the clear case for manually picking the arrival time of the microseismic event due to the confinement of the seismic energy to a dominant region (shown in detail by the rectangle and its zoomed version). We pick the first-break of the microseismic event on trace 1 and save it as $n_r= 289$ samples.
 In most cases, manual picking of the arrival time of the microseismic event can be done easily with the TFT of the first trace. For cases with even lower PSNR, when the visibility of the signal is not clear in time-frequency domain, more than one trace can be transformed to the time-frequency domain to make a robust decision.

\begin{figure*}
\begin{center}
\includegraphics[width=1.0\textwidth]{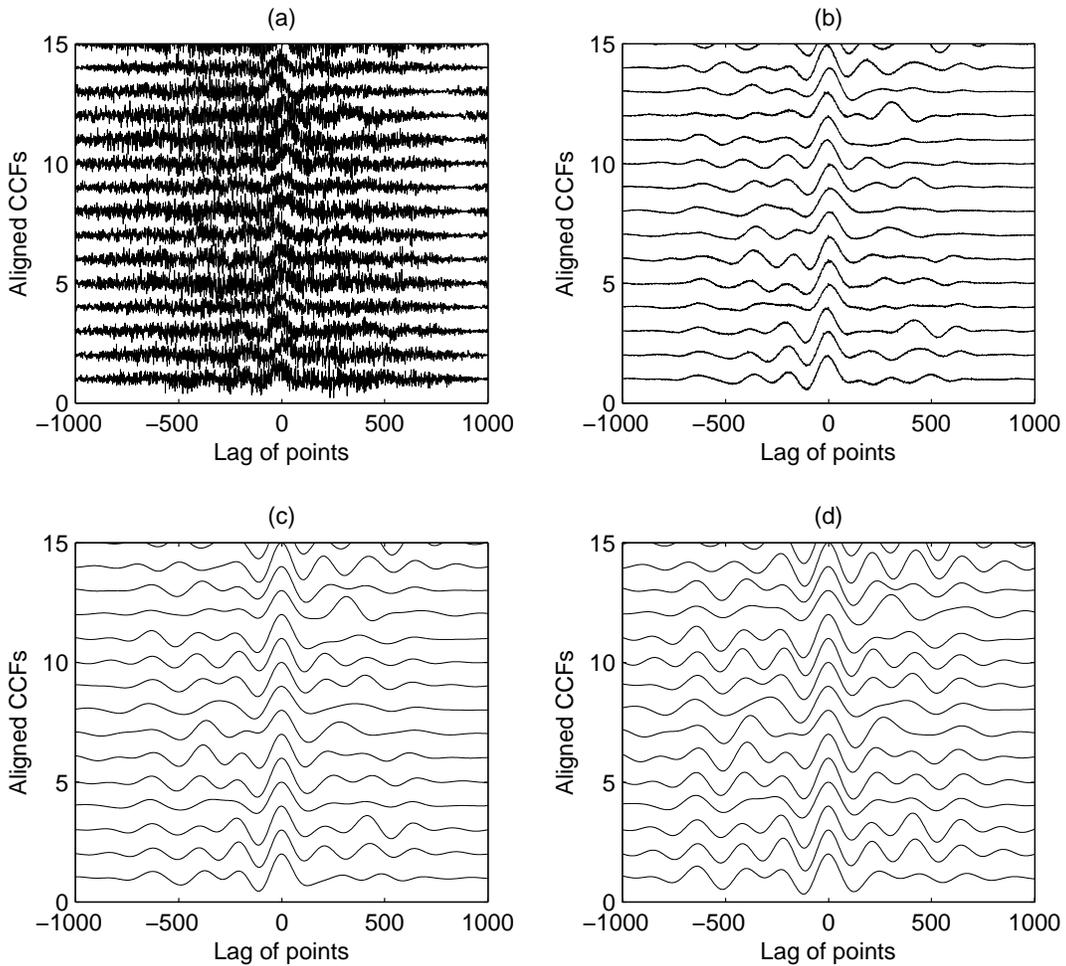}
\caption{\small CCFs after alignment for (a) iteration 0, (b) iteration 1, (c) iteration 2, and (d) iteration 3.}
\label{fig:CC_syn}
\end{center}
\end{figure*}

 Next, the iteration is started by performing all pairwise cross-correlations. Raw cross-correlations are not shown here since they do not give any useful information to the reader; instead aligned CCFs are shown in Figure \ref{fig:CC_syn}a (and to make the picture clear only 15 out of 91 CCFs are shown). The alignment of CCFs is obtained by shifting each cross-correlation such that its maximum value occurs at lag $\tau=0$.
 In succeeding iterations, modified CCFs are produced in the update of Step 4, so Figure \ref{fig:CC_syn}b, \ref{fig:CC_syn}c, and \ref{fig:CC_syn}d also depict the aligned modified cross-correlations for each iteration. It is clear that the aligned modified CCFs after iteration 1 are less noisy when compared to the aligned CCFs after iteration 0. One notable feature that can also be seen in these figures is that, as the number of iterations increases, the modified CCFs exhibit an oscillating behavior.
%% {\color{red} (DELETE: The reason is that as we iterate, correlation between data samples increases and, therefore, the noise and signal also become correlated. Hence, the basic assumption that noise and signal are uncorrelated is no longer valid. )  Is this really true?}
  The reason is that as we iterate, the Fourier transform of the modified CCFs is essentially the Fourier transform of the signal raised to higher powers. As a result the dominant frequency amplitudes will be enhanced multiple times, and finally yields modified cross-correlations that are very close to a cosine function. Therefore, the stacked cross-correlation after a number of iterations does not help to locate an accurate first-break, because of its oscillatory behavior.

\begin{figure}[htbp]
\begin{center}
\includegraphics[width=0.75\textwidth]{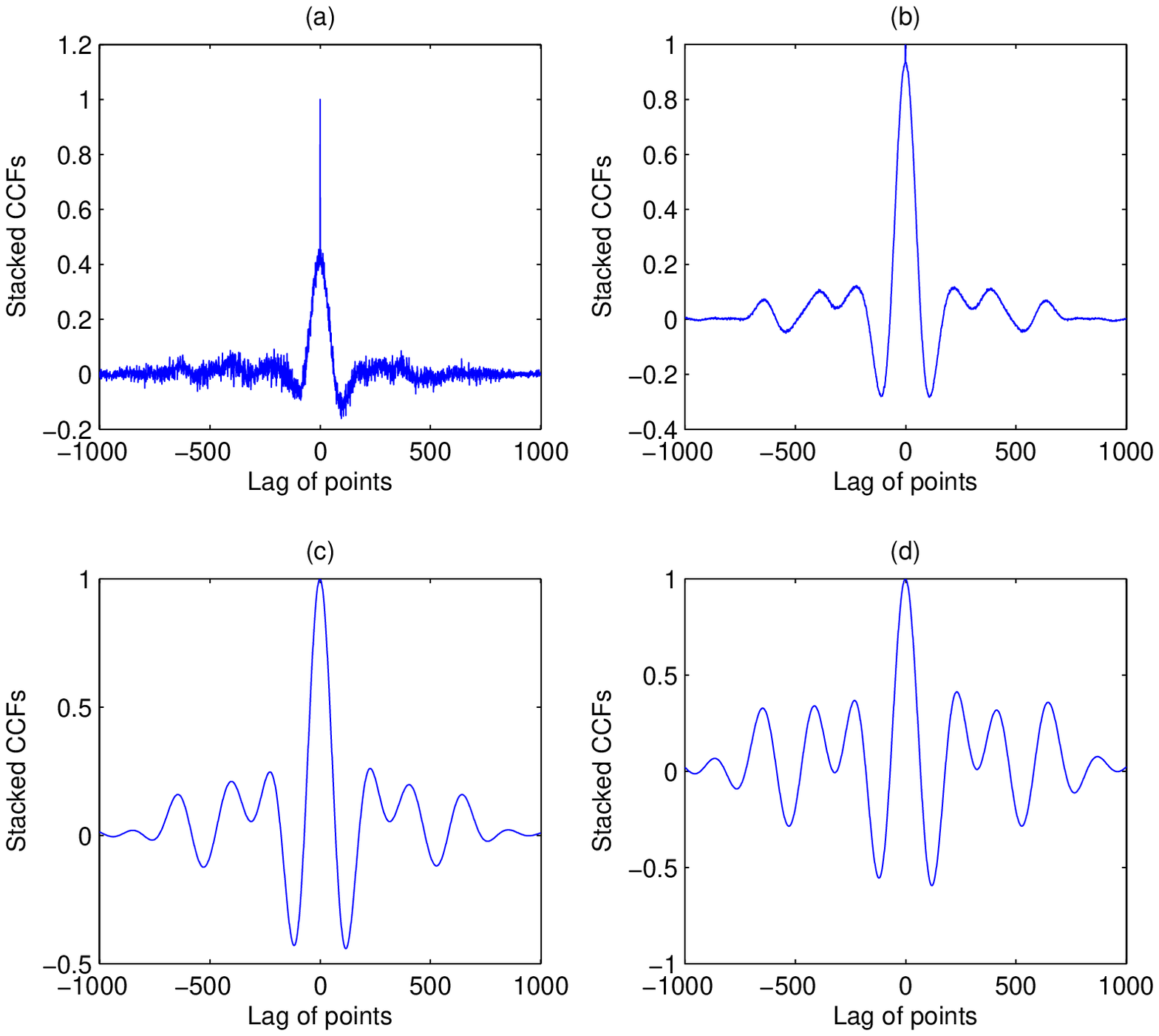}
\caption{Aligned CCFs after stacking for (a) iteration 0, (b) iteration 1, (c) iteration 2, and (d) iteration 3.}
\label{fig:SCC}
\end{center}
\end{figure}

   After alignment the (modified) cross-correlations are stacked with the result being shown in Figure \ref{fig:SCC}. This figure validates the points highlighted for Figure \ref{fig:CC_syn}. The spike in the stacked cross-correlation after iteration 0 is due to the noise and misalignment of the CCFs.
  These problems are reduced significantly and removed automatically after iteration 1 because of increasing signal contribution and decreasing noise contribution with iteration number.

%  \begin{figure*}
%         \centering
%         \begin{subfigure}[b]{0.475\textwidth}
%             \centering
%             \includegraphics[width=\textwidth]{Fig/CC_syn_0}
%             \end{subfigure}
%         \hfill
%         \begin{subfigure}[b]{0.475\textwidth}
%             \centering
%             \includegraphics[width=\textwidth]{Fig/CC_syn_1}
%            \end{subfigure}
%         \vskip\baselineskip
%         \begin{subfigure}[b]{0.475\textwidth}
%             \centering
%             \includegraphics[width=\textwidth]{Fig/CC_syn_2}
%            \end{subfigure}
%         \quad
%         \begin{subfigure}[b]{0.475\textwidth}
%             \centering
%             \includegraphics[width=\textwidth]{Fig/CC_syn_3}
%             \end{subfigure}
%         \caption[]
%         {\small CCFs after alignment for (a) iteration 0, (b) iteration %   1, (c) iteration 2, and (d) iteration 3.}
%         \label{fig:CC_syn}
%     \end{figure*}

\begin{figure}[hbtp]
\begin{center}
\includegraphics[width=0.99\textwidth]{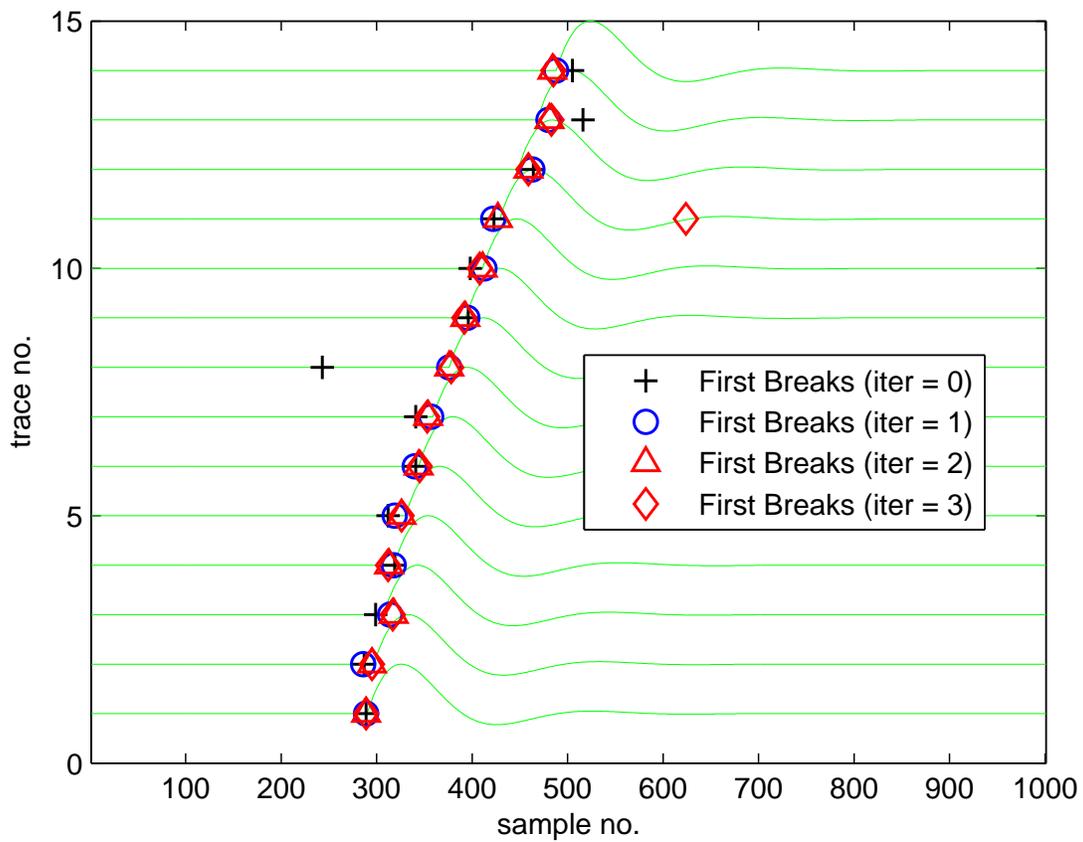}
\caption{Picked first breaks using minimum-phase wavelet (dominant frequency of 5Hz) of various iterations. SNR = $-$12\,dB}
\label{fig:PFB}
\end{center}
\end{figure}

        \begin{figure}[htbp]
\begin{center}
\includegraphics[width=1\textwidth]{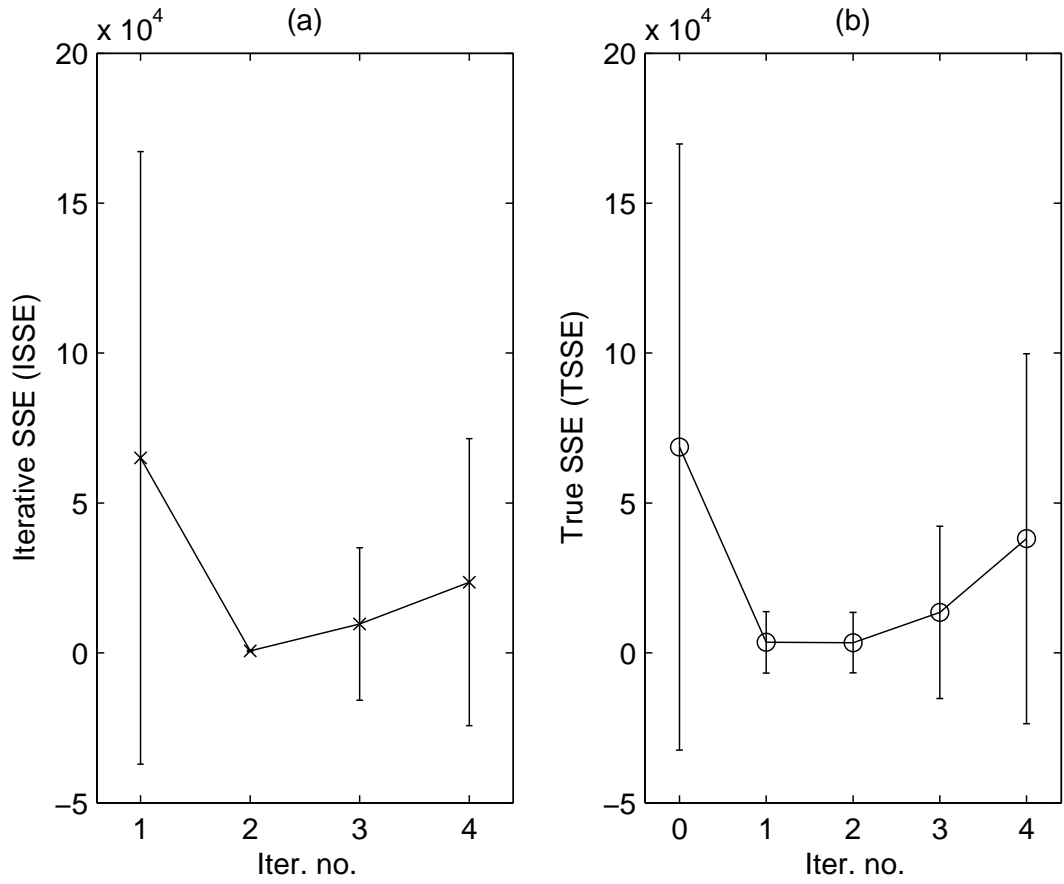}
\caption{Results of Monte Carlo test (with error bars) on noisy synthetic data with 20 trials for (a) ISSE and (b) TSSE.}
\label{fig:SSE}
\end{center}
\end{figure}

\begin{figure}[htbp]
        \begin{center}
                \includegraphics[width=0.6\textwidth]{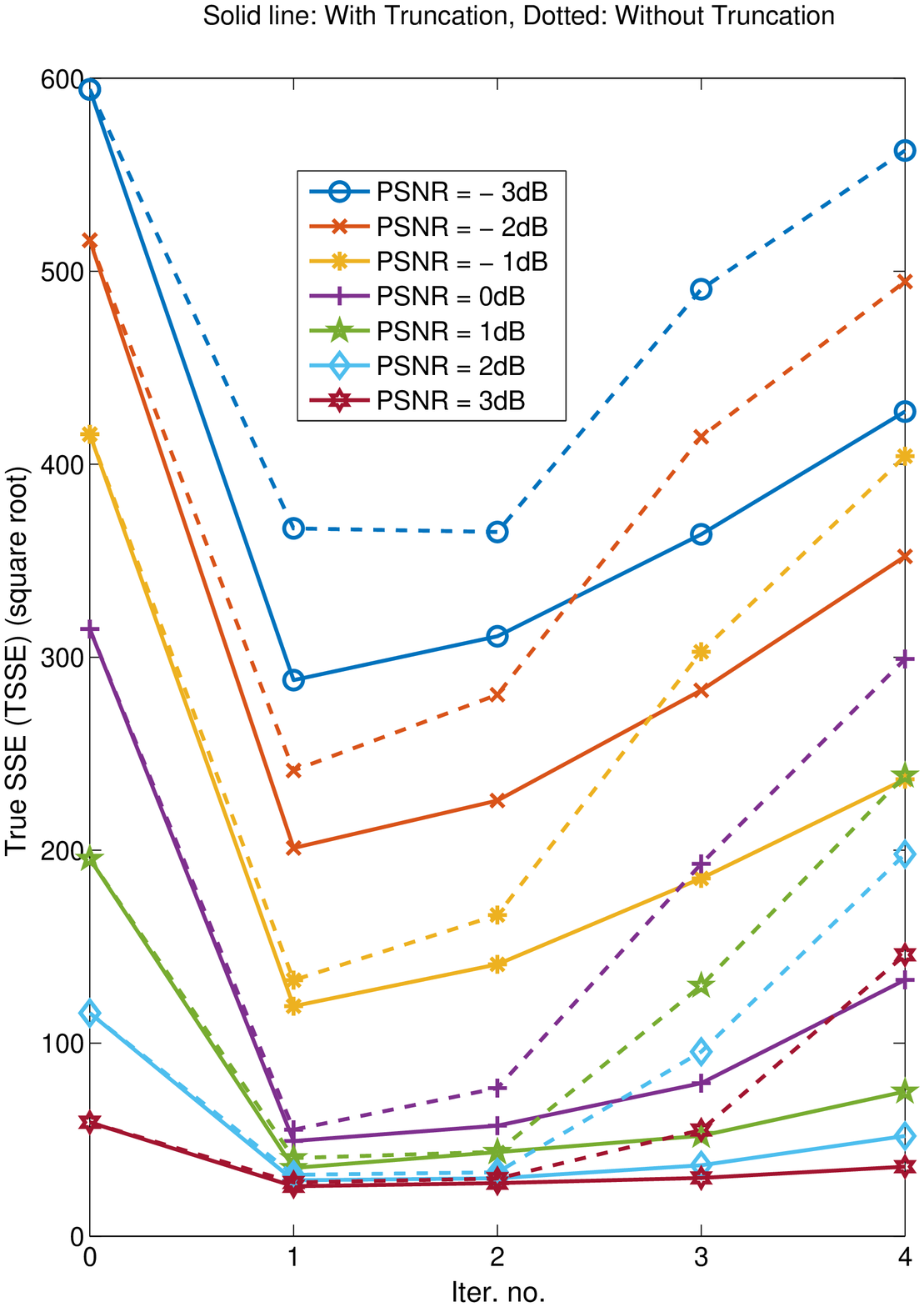}
\caption{Square root of TSSE at different PSNRs from Monte Carlo testing on noisy synthetic data with 20 trials. Solid lines represent results with truncation to [-350,350], whereas, dotted lines represent results without truncation}
                \label{fig:SSESNR}
        \end{center}
\end{figure}

The raw traces after posting the first breaks from each iteration are shown in Figure \ref{fig:PFB}. Since we have to iterate through step $1$ and step $2$ of the proposed algorithm, therefore, we  use the iterative sum of squared errors (ISSE) as a stopping criteria.
 ISSE is calculated as the sum of the difference of first breaks obtained for two consecutive iterations (current and previous)
 \begin{equation}
    \text{ISSE}(i) = \sum\limits_{m=1}^M \left(n^i_m - n^{i-1}_m \right)^2
 \end{equation}
We also compute the true sum of squares error (TSSE) as the sum of the differences of first breaks and the true first breaks
\begin{equation}
   \text{TSSE}(i) =  \sum\limits_{m=1}^M \left(n^i_m - n_{\text{true}} \right)^2
\end{equation}

Plots of the ISSE and TSSE versus iteration number are depicted in Figure \ref{fig:SSE}.
For plotting ISSE/TSSE, 20 Monte Carlo simulations are performed.
 The behavior of TSSE confirms that we can base our stopping criteria on the ISSE. It can be seen that TSSE first decreases and then increases after iteration 2 and ISSE follows the same pattern. The increase in TSSE is due to the enhanced dominant frequency  amplitudes mentioned earlier for Figures \ref{fig:CC_syn} and \ref{fig:SCC}.

The performance of the algorithm for different PSNRs and iterations is shown in Figure \ref{fig:SSESNR}. A 5Hz minimum-phase Ricker wavelet is used as the source waveform and 500 Monte Carlo simulations were performed. Figure \ref{fig:SSESNR} shows that the TSSE increases after first iteration for all PSNRs. Figure \ref{fig:SSESNR} also compares   the performance of the proposed technique with and without truncation window $T[\tau]$ (represented by solid lines and dotted line, respectively). With truncation, the cross-correlations in step 4 (see Table \ref{alg:iter}) of the algorithm are taken as zero after $N_T$ (or $-N_T$ on the negative lag) samples. The value of $N_T$ = 350 is used in the simulation, which is quarter of the total signal duration (1000 samples). On the other hand, without truncation cross-correlations are left unchanged. The performance improvement due to truncation window $T[\tau]$ is  evident from the figure.

\begin{figure}[hbtp]
\begin{center}
\includegraphics[width=0.99\textwidth]{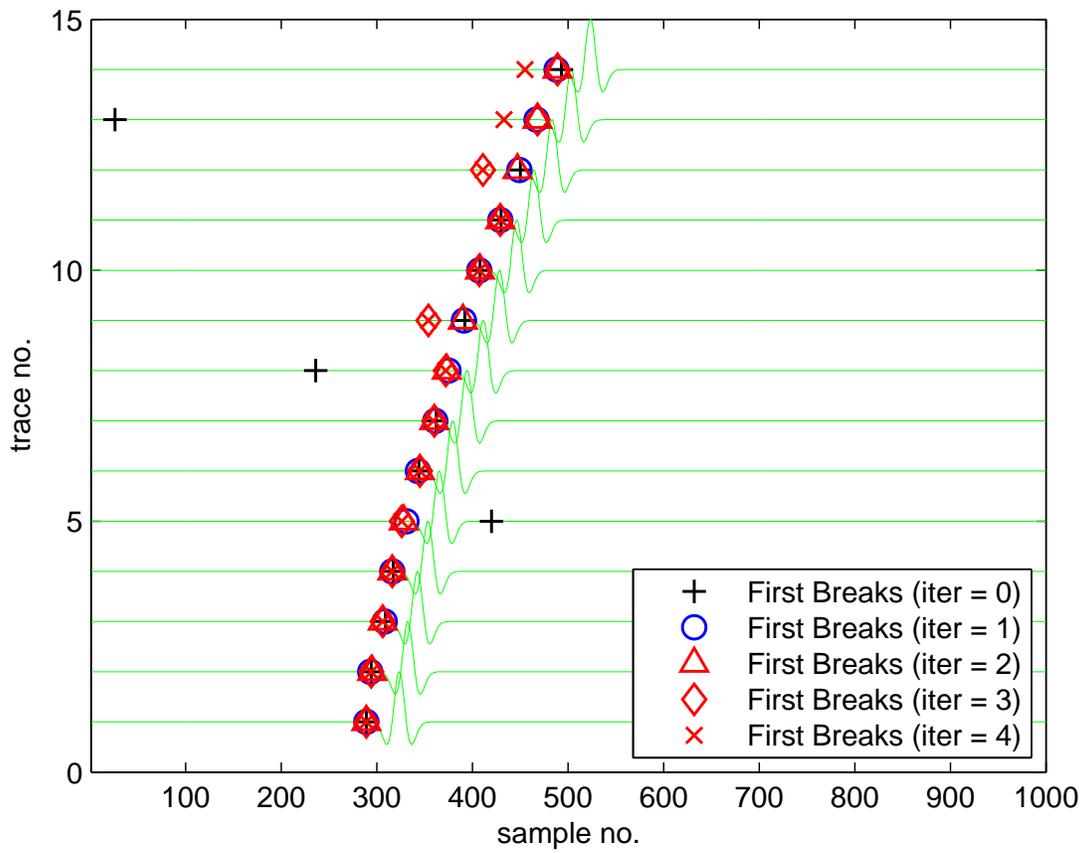}
\caption{Picked first breaks using Ricker wavelet (dominant frequency of 30Hz) of various iterations.  SNR = $-$12\,dB}
\label{fig:rfb}
\end{center}
\end{figure}

\begin{figure}[htbp]
        \begin{center}
                \includegraphics[width=0.99\textwidth=0.6\textwidth]{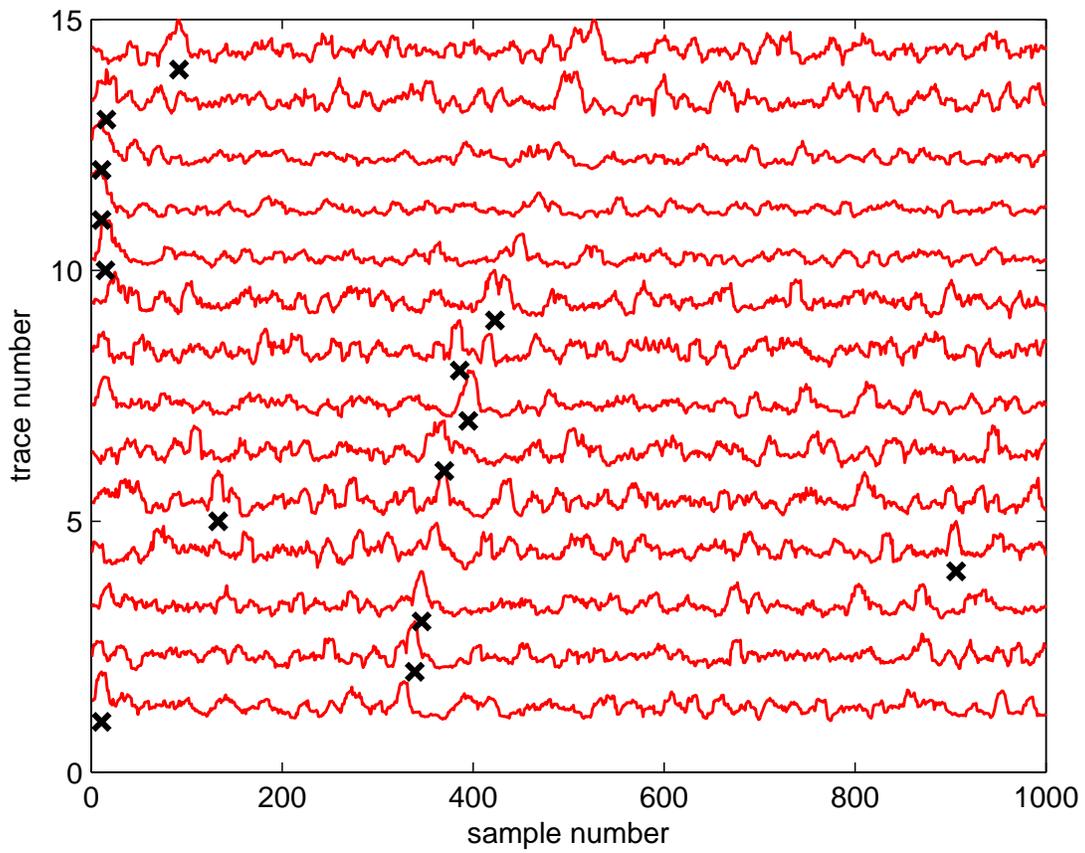}
\caption{ STA/LTA ratio with zero-phase Ricker wavelet of 30Hz and SNR = $-$12\,dB. The lengths of the STA and LTA windows are set to 10 and 50 samples, respectively.}
                \label{fig:STA}
        \end{center}
\end{figure}

 We also apply our proposed algorithm to zero-phase Ricker wavelet with dominant frequency of 30Hz. Figure \ref{fig:rfb} shows the algorithm converges in first iteration.
  In order to show the effectiveness of the proposed method, we also present the result of the STA/LTA method (with zero-phase Ricker wavelet of 30Hz) in Figure \ref{fig:STA}. It is apparent from the figure that  the STA/LTA method is not applicable for low SNR signal, since many false peaks are comparable or even surpass the local peaks on the correct arrivals.
\clearpage

\subsection{Test on real data}
\begin{figure*}
        \centering
        \begin{subfigure}[b]{0.475\textwidth}
            \centering
            \includegraphics[width=1\textwidth]{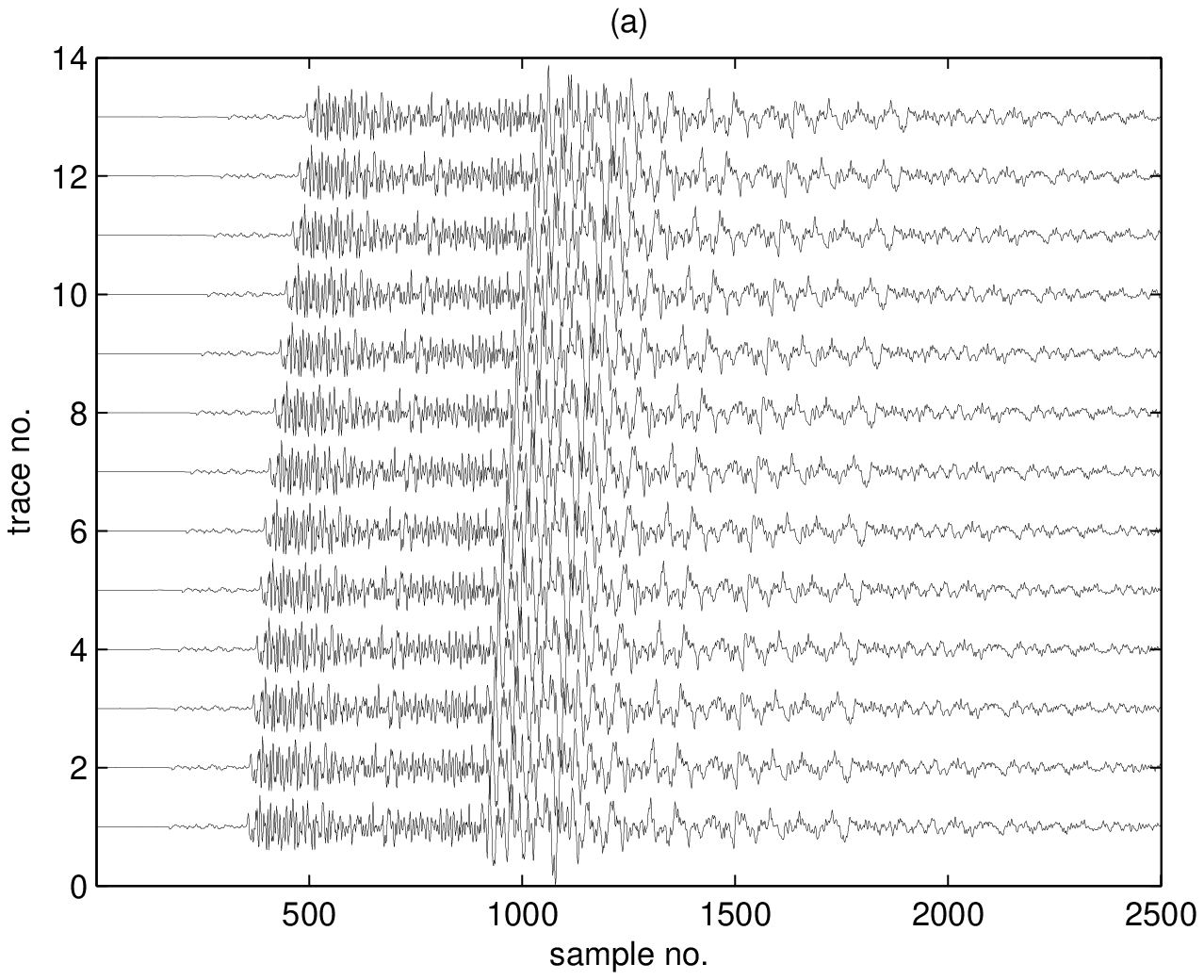}
        \end{subfigure}
        \hfill
        \begin{subfigure}[b]{0.475\textwidth}
            \centering
            \includegraphics[width=1\textwidth]{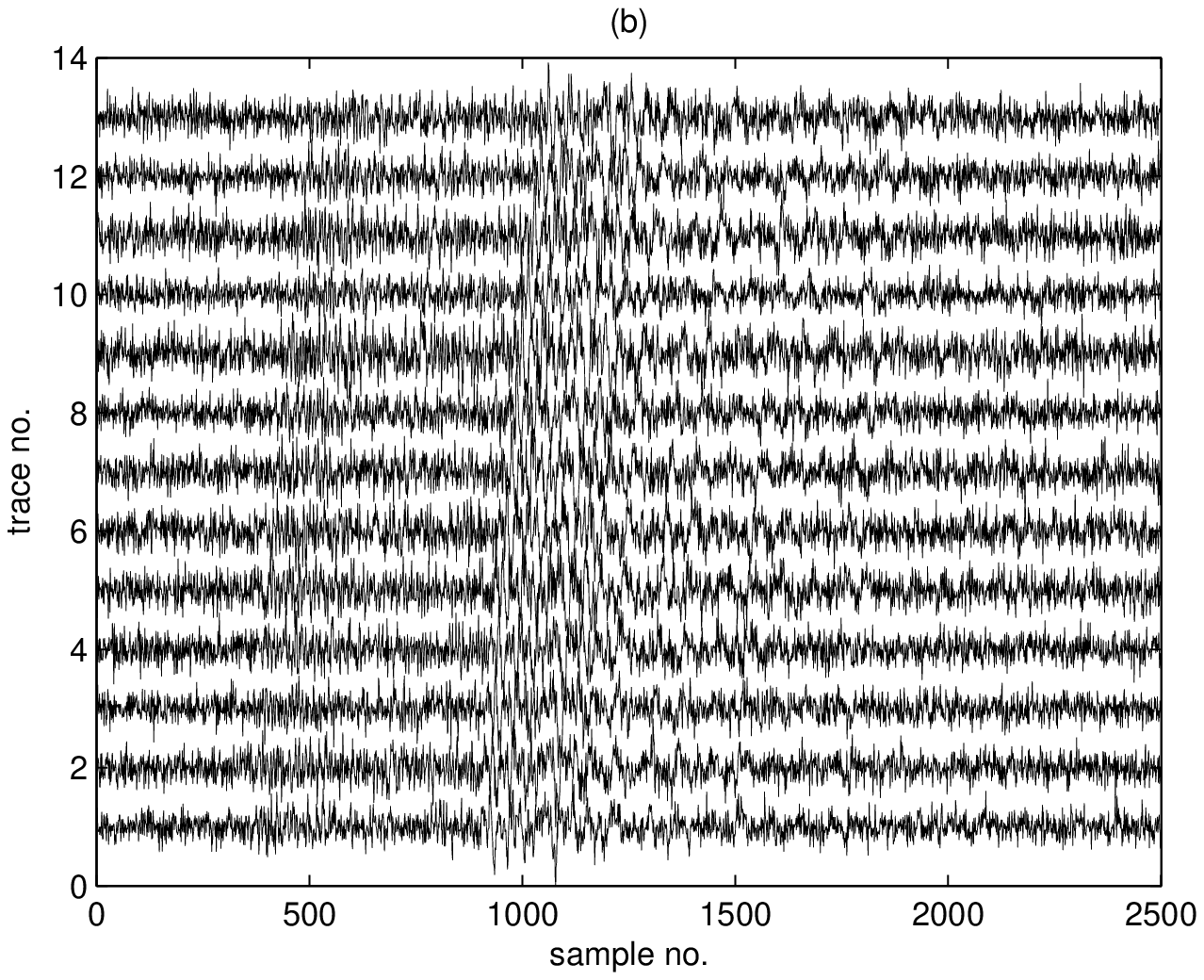}
        \end{subfigure}
        \caption
        {\small Real data; (a) noiseless traces; (b) noisy traces at PSNR = 8 dB.}
      \label{fig:noisy_real}
    \end{figure*}

The algorithm was originally tested on the real data set from the local industry in the Middle East. However, due to proprietary issues, we are unable to include those results. In order to demonstrate the effectiveness of the proposed method, we now apply our work-flow on a field data set (Figure \ref{fig:noisy_real}a). The data used in this study comes from the
High Resolution Seismic Network (HRSN) operated by Berkeley Seismological Laboratory,
University of California, Berkeley, and the Northern California Seismic Network
(NCSN) operated by the U.S. Geological Survey, Menlo Park, and is distributed by
the Northern California Earthquake Data Center (NCEDC). The sampling frequency
is 250 Hz. We repeat a single waveform 13 times with time shifts that simulate multiple traces recorded at distances far from the source. The microseismic event originally has a good SNR and did not need first-break enhancement; but for testing we add zero-mean white Gaussian noise. The PSNR is set to 8\,dB in order to make the first-break challenging for automatic picking, as shown in  Figure \ref{fig:noisy_real}b.

\begin{figure}[htbp]
\begin{center}
\includegraphics[width=0.9\textwidth]{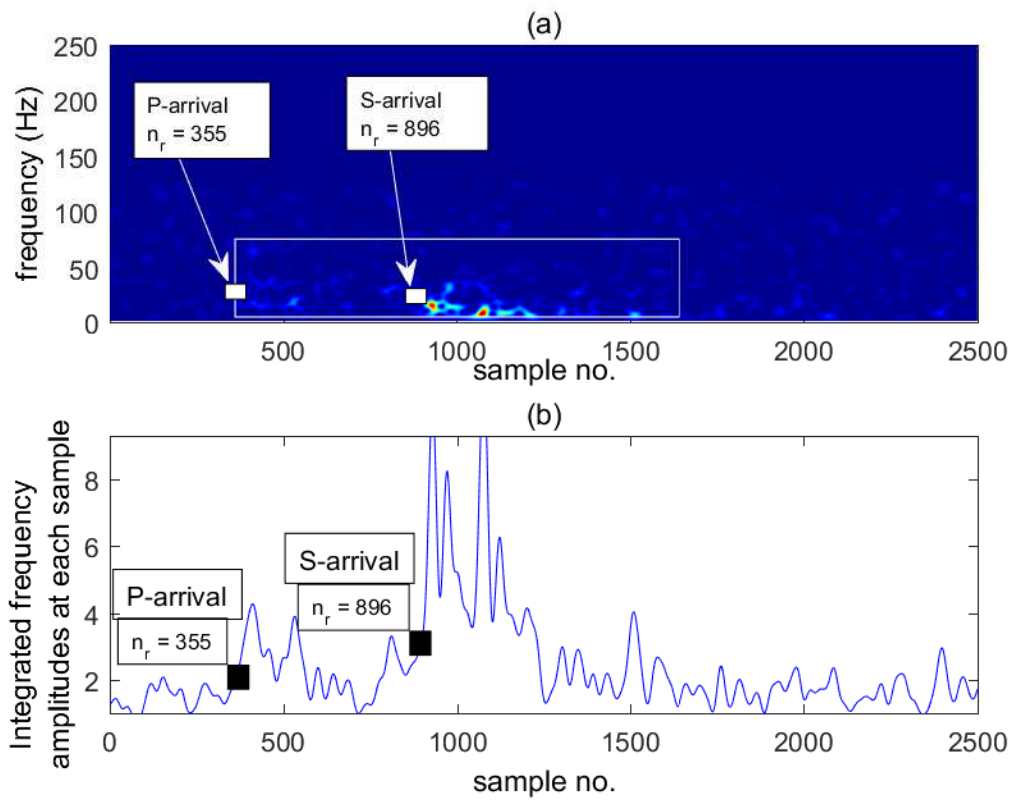}
\caption{(a) Spectrogram of the first trace and picked first break (indicated by the small square) and (b) sum of magnitudes across frequency at each time sample.}
\label{fig:TF_real}
\end{center}
\end{figure}

In Figure \ref{fig:TF_real}a, we examine the TFT computed from the real data. It appears that trace 1 gives a good indication of the arrival time  of P- and S-wave of the microseismic event in the time-frequency representation as depicted in Figure \ref{fig:TF_real}a. Therefore, trace 1 was used as the reference trace to manually pick the arrival time of the P-wave  (or S-wave) microseismic event and saved as $n_r  = 355$ (or $ = 896$)  samples. It should be noted here that first, we pick the S-arrival reference and apply the aforementioned algorithm to get the S-arrivals of all the traces. Next, P-arrival  reference is picked and algorithm is again applied after removing the S-waves. This gives the P-arrivals of all the traces. In Figure \ref{fig:TF_real}b we show cumulative frequency amplitude in each time sample, which serves as a good reference for the time picking in the spectrogram.

\begin{figure*}
\begin{center}
\includegraphics[width=1.0\textwidth]{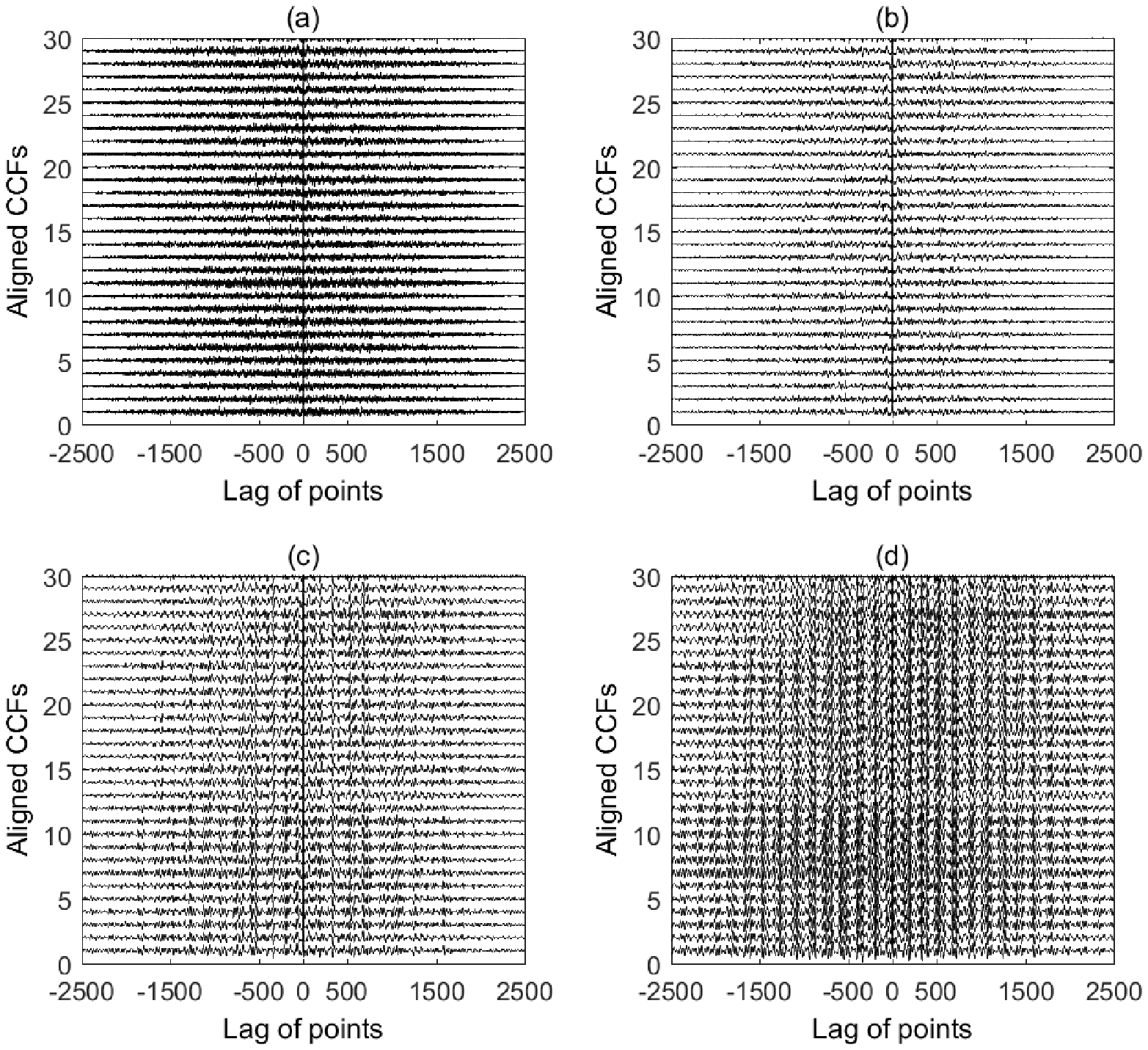}
\caption{
 CCFs after alignment for (a) iteration 0, (b) iteration 1, (c) iteration 2, and (d) iteration 3.}
\label{fig:CC_real}
\end{center}
\end{figure*}

%  \begin{figure*}
%         \centering
%         \begin{subfigure}[b]{0.475\textwidth}
%             \centering
%             \includegraphics[width=\textwidth]{Fig/CC_real_0}
%            \end{subfigure}
%         \hfill
%         \begin{subfigure}[b]{0.475\textwidth}
%             \centering
%             \includegraphics[width=\textwidth]{Fig/CC_real_1}
%            \end{subfigure}
%         \vskip\baselineskip
%         \begin{subfigure}[b]{0.475\textwidth}
%             \centering
%             \includegraphics[width=\textwidth]{Fig/CC_real_2}
%           \end{subfigure}
%         \quad
%         \begin{subfigure}[b]{0.475\textwidth}
%             \centering
%             \includegraphics[width=\textwidth]{Fig/CC_real_3}
%           \end{subfigure}
%         \caption[]
%         {\small CCFs after alignment for (a) iteration 0, (b) iteration 1, (c) iteration 2, and (d) iteration 3.}
%         \label{fig:CC_real}
%     \end{figure*}
%

\begin{figure}[htbp]
\begin{center}
\includegraphics[width=0.75\textwidth]{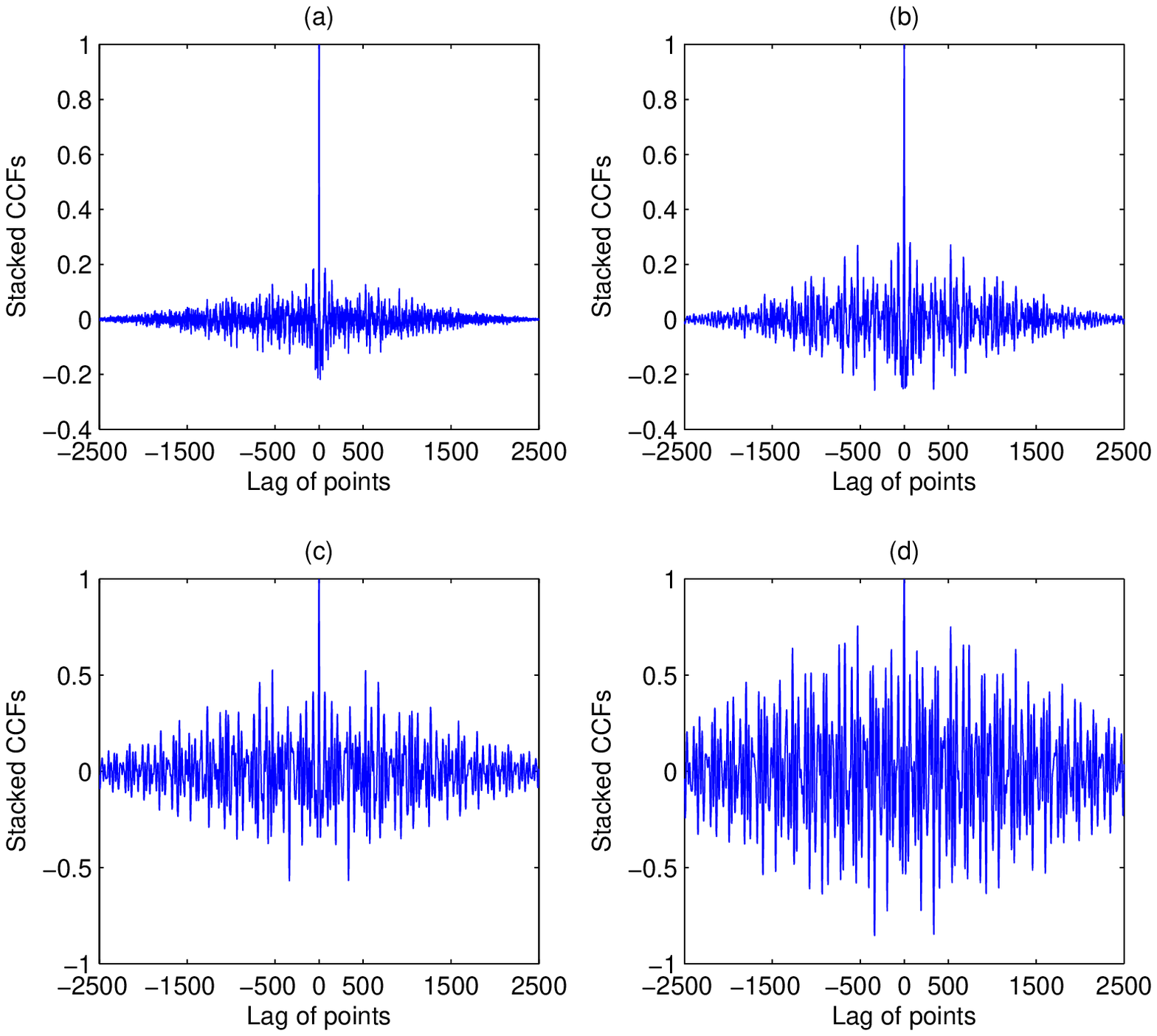}
\caption{Aligned CCFs after stacking for (a) iteration 0, (b) iteration 1, (c) iteration 2, and (d) iteration 3.}
\label{fig:stackedCCF_real}
\end{center}
\end{figure}

\begin{figure}[htbp]
\begin{center}
\includegraphics[width=0.99\textwidth]{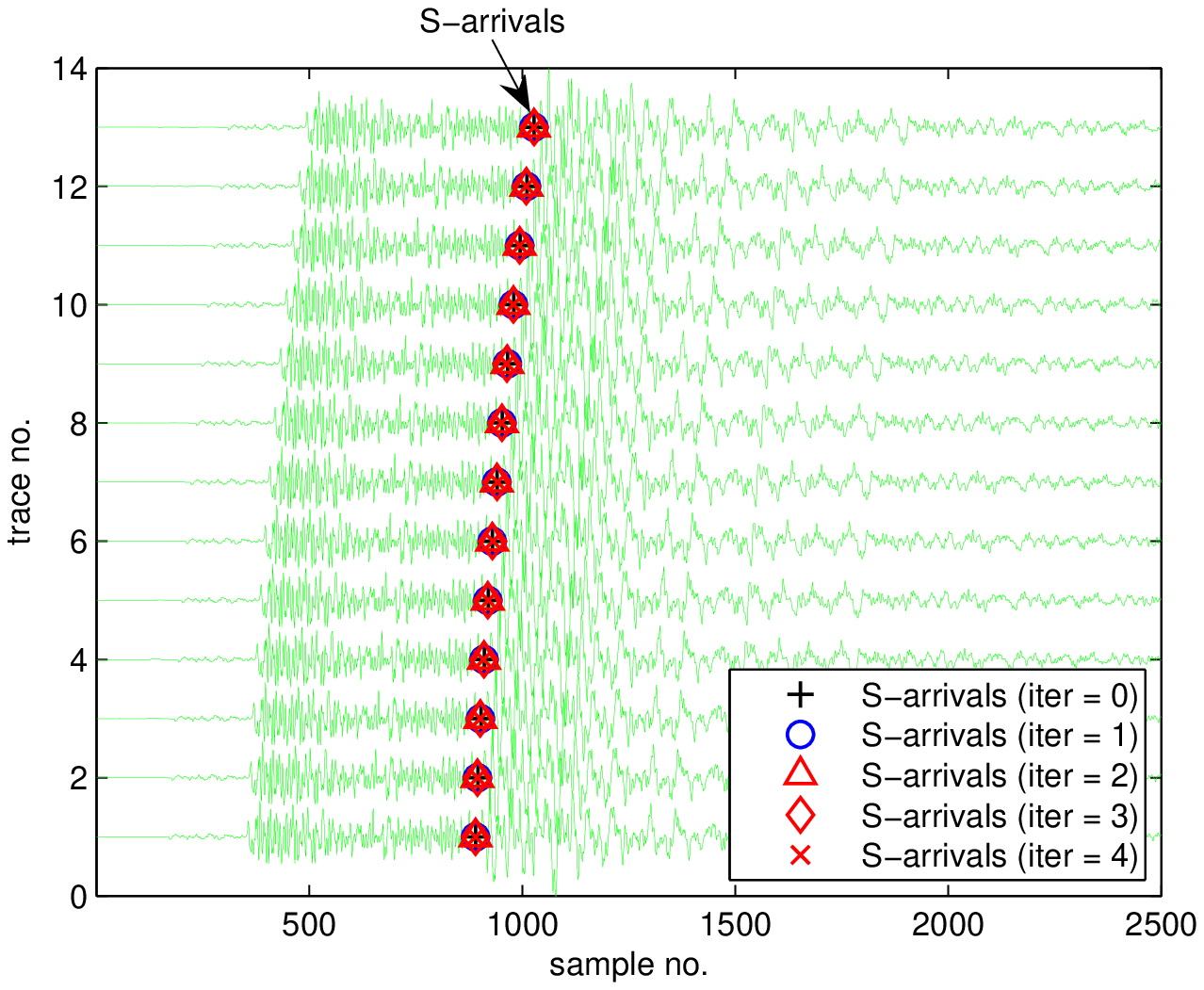}
\caption{Picked S-arrivals of various iterations.}
\label{fig:firstbreak_real_S}
\end{center}
\end{figure}

\begin{figure}[htbp]
\begin{center}
\includegraphics[width=1\textwidth]{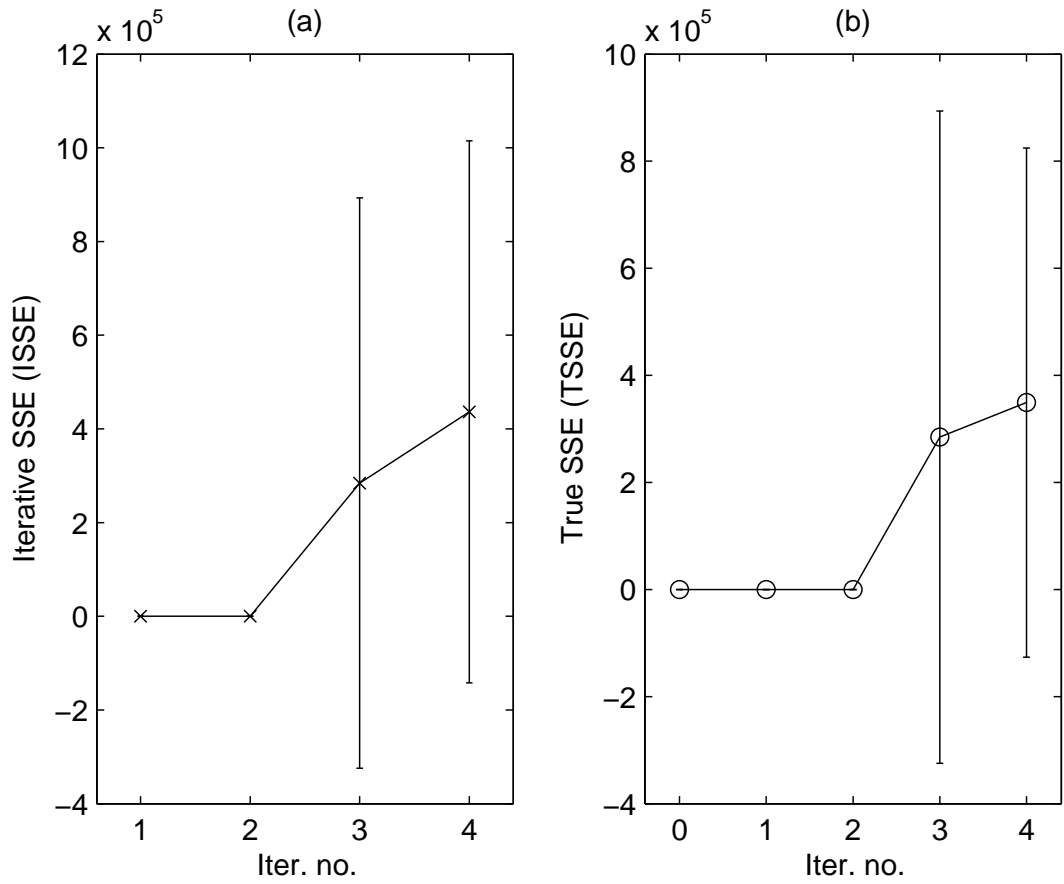}
\caption{Results of Monte Carlo test on noisy real data with 20 trials for (a) ISSE and (b) TSSE.}
\label{fig:SSE_real}
\end{center}
\end{figure}

\begin{figure}[htbp]
\begin{center}
\includegraphics[width=0.99\textwidth]{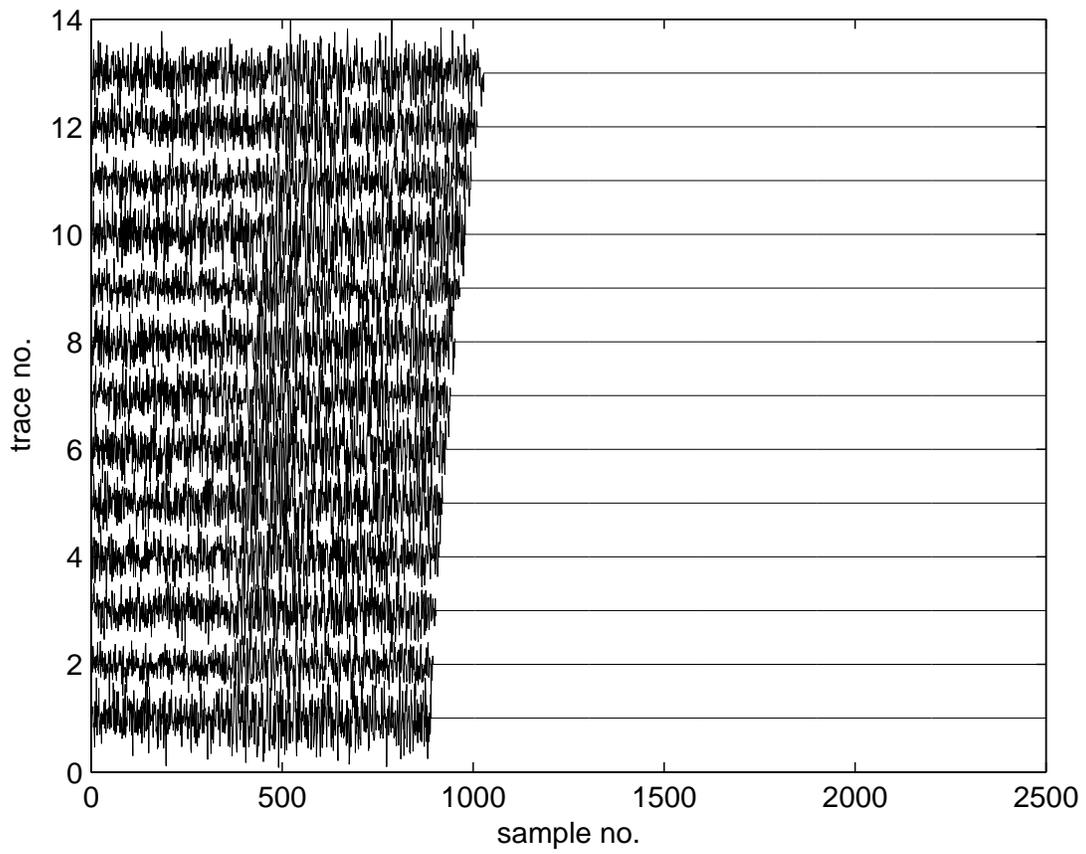}
\caption{Noisy real data set with only P-waves.}
\label{fig:firstbreak_real_S_mute}
\end{center}
\end{figure}

\begin{figure}[htbp]
\begin{center}
\includegraphics[width=0.99\textwidth]{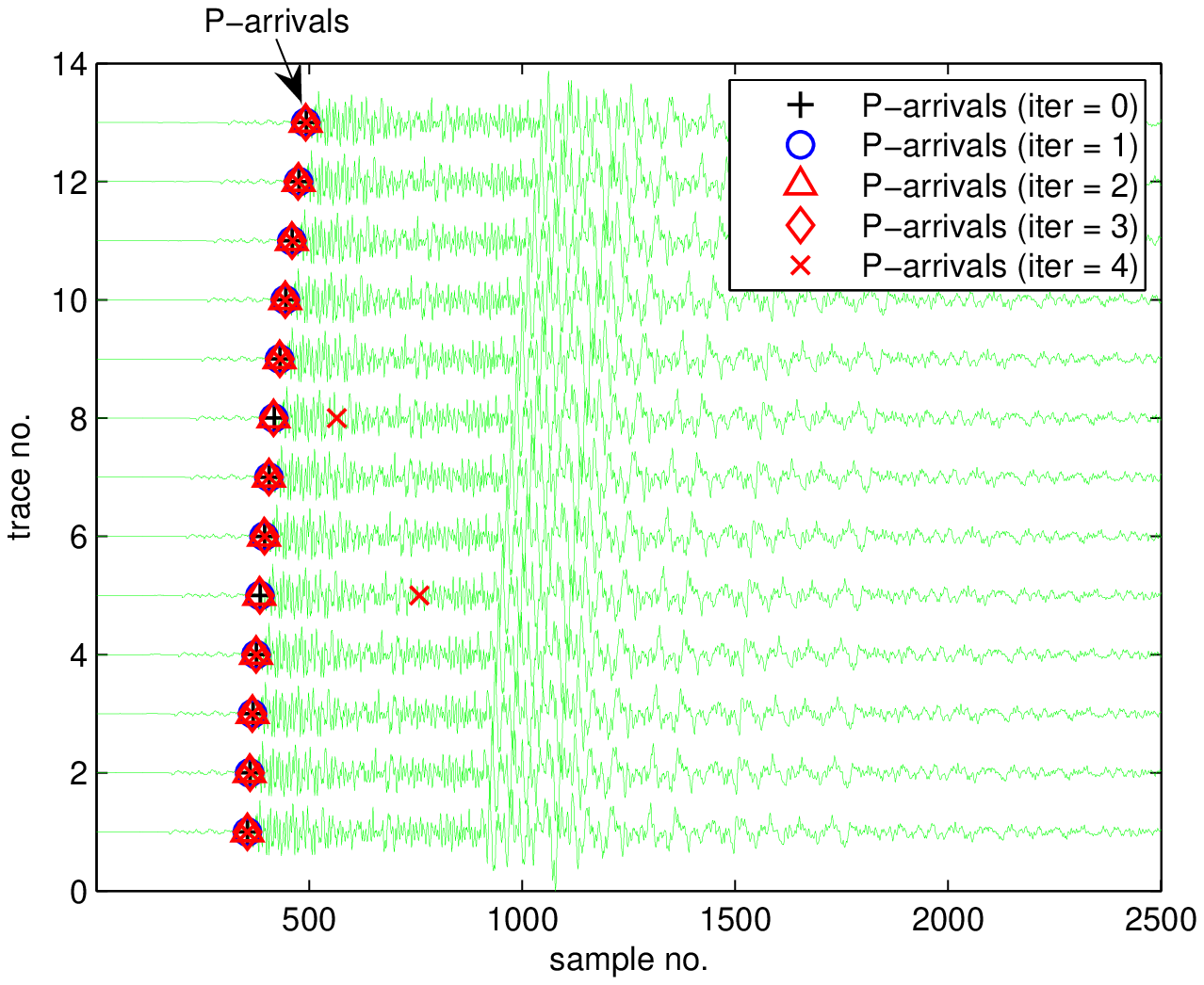}
\caption{Picked P-arrivals of various iterations.}
\label{fig:firstbreak_real_P}
\end{center}
\end{figure}

Figures \ref{fig:CC_real} and \ref{fig:stackedCCF_real} show the  CCFs after alignment and stacking, respectively, during the iterative algorithm of our proposed approach, respectively. The figures for CCFs  and stacked CCFs are plotted for S-arrivals only, however, for P-arrivals same procedure should be followed. The raw traces after posting the S-arrivals from each iteration are depicted in Figure \ref{fig:SSE_real}, as well as plots of the ISSE and TSSE versus iteration number in Figure \ref{fig:firstbreak_real_S} (resulting from 20 Monte Carlo simulations).
Next, to find P-arrivals, S-waves are muted according to Eqs. (\ref{win1}) and (\ref{win2}) and the P-waves only data set is shown in figure \ref{fig:firstbreak_real_S_mute}. After application of the algorithm again, P-arrivals are shown in Figure \ref{fig:firstbreak_real_P}, which confirms the algorithm's performance for P-arrivals as well. It can be seen from these figures that results of testing the proposed algorithm on real data agree well with tests on synthetic.
\section{Discussion and conclusion}
We have developed a new approach for picking first breaks of noisy microseismic records using an iterative approach based on seismic interferometry. Unlike existing interferometry-based approaches our approach does not involve convolution of the stacked CCF with raw traces, which ensures that the raw data does not mix with the enhanced stacked record.
Using synthetic data as well as real data, we demonstrate a significant improvement in first break picking. Consequently, the enhanced first breaks will improve microseismic event localization over the reservoir.
Results from synthetic data show that the SSE between the true and estimated first breaks does not change considerably after the first iteration, which indicates the fast convergence rate of the proposed method. Therefore, we recommend  iterating only once and  using more iterations only if the SSE changes slowly. The reason for this observation is that in the first CCF between the raw traces the signal could be highly variable from trace to trace due to effects of noise and attenuation. In contrast, in the first iteration, we cross-correlate the aligned stacked CCF which essentially consists of an average auto-correlation of the signals with the cross-correlations of trace pairs which consists of an approximate auto-correlation of the signal in this trace pair; hence, they should exhibit more similarity than raw traces.
We believe that the approach proposed in this study is important and useful for the petroleum industry as well as academic and research institutions interested in continuous reservoir monitoring using microseismic data. Thus, we recommend using our newly developed workflow as a first-break picking tool during the processing of microseismic data recorded over the reservoir.

\section{Acknowledgements}

We thank King Fahd University of Petroleum \& Minerals (KFUPM) for supporting this study through the Center for Energy and Geo Processing (CeGP) at KFUPM under grant number GTEC1311.

\bibliographystyle{apalike}
\bibliography{refs}

\end{document}